\documentclass[a4paper,11pt]{article}

\usepackage{amsmath,amssymb}
\usepackage{bm}
\usepackage{graphicx}
\usepackage{authblk}
\usepackage{natbib}
\setcitestyle{authoryear,open={(},close={)}}
\usepackage[top=3cm,bottom=3cm,left=2.5cm,right=2.5cm]{geometry}

\title{
An efficient branch-and-bound algorithm\\ for submodular function maximization
}

\author[1,2]{Naoya Uematsu \footnote{\texttt{naoya.uematsu@riken.jp}}}
\author[1,2]{Shunji Umetani \footnote{\texttt{umetani@ist.osaka-u.ac.jp}}}
\author[1,3]{Yoshinobu Kawahara \footnote{\texttt{yoshinobu.kawahara@riken.jp}}}
\affil[1]{RIKEN Center for Advanced Intelligence Project, Quantative Biology Center, 6-2-4 Fruedai, Suita, Osaka, 565-0874, Japan.}
\affil[2]{Graduate School of Information Science and Technology, Osaka University, 1-5 Yamadaoka, Suita, Osaka, 565-0871, Japan.}
\affil[3]{The Institute of Scientific and Industrial Research, Osaka University, 8-1 Mihogaoka, Ibaraki, Osaka, 567-0047, Japan.}

\date{}

\begin{document}
\maketitle

\begin{abstract}
The submodular function maximization is an attractive optimization model that appears in many real applications.
Although a variety of greedy algorithms quickly find good feasible solutions for many instances while guaranteeing $(1-1/e)$-approximation ratio, we still encounter many real applications that ask optimal or better feasible solutions within reasonable computation time.
In this paper, we present an efficient branch-and-bound algorithm for the non-decreasing submodular function maximization problem based on its binary integer programming (BIP) formulation with a huge number of constraints.
Nemhauser and Wolsey developed an exact algorithm called the constraint generation algorithm that starts from a reduced BIP problem with a small subset of constraints taken from the constraints and repeats solving a reduced BIP problem while adding a new constraint at each iteration.
However, their algorithm is still computationally expensive due to many reduced BIP problems to be solved.
To overcome this, we propose an improved constraint generation algorithm to add a promising set of constraints at each iteration.
We incorporate it into a branch-and-bound algorithm to attain good upper bounds while solving a smaller number of reduced BIP problems.
According to computational results for well-known benchmark instances, our algorithm achieved better performance than the state-of-the-art exact algorithms.
\end{abstract}

\section{Introduction}
There are many problems in computer science that are formulated as the submodular function maximization, such as active learning~\citep{Krause2014}, sensor placement~\citep{Golovin2011,Kawahara2009,Kratica2001}, influence spread~\citep{Kempe2003} and feature selection~\citep{Jovic2015,Yu2004,Das2011}. 
Submodular function is a set function $f$ satisfying $f(S) + f(T) \geq f(S \ \cap \ T ) + f(S\ \cup \ T)$ for all $S, T \subseteq N$, where $N$ is a finite set. 
Submodular functions can be considered as discrete counterparts of convex functions through the continuous relaxation called the Lovasz extension \citep{Lovasz1983}. 
We address the problem of maximizing a submodular function with a cardinality constraint, formulated as
\begin{equation}
\label{eq:subm}
\begin{array}{lll}
\textnormal{maximize} & f(S) \\
\textnormal{subject to} & |S| \leq k, &  S \subseteq N,
\end{array}
\end{equation}
where $N = \{1, \dots , n\}$, $n$ is the size of the finite set and $k \leq n$ is a positive integer comprising the cardinality constraint. 
This paper focuses on non-decreasing submodular functions. 
A submodular function is non-decreasing if $f(S) \leq f(T)$ for all $S \subseteq T$ and $f(\emptyset) = 0$.
This problem is well known as NP-hard, unlike submodular function minimization~\citep{Lee2009}. 

For the problem~(\ref{eq:subm}), \citet{Nemhauser1978} invoked a greedy algorithm which achieves an approximation ratio $(1-1/e)$ $(\approx 0.63)$.
For large-scale instances, the greedy algorithm is not efficient since it takes $\mathrm{O}(nk)$ oracle queries.
To overcome this, \citet{Minoux1978} improved the performance of the greedy algorithm in practice.

Although the greedy algorithm quickly finds good feasible solutions for many instances of submodular maximization, we often encounter real applications that ask optimal or better feasible solutions within reasonable computation time.
For instance, the feature selection problem asks to select the essential features to represent a model with minimal loss of information.
The greedy algorithm often fails to select essential elements that can not be removed without affecting the original conditional target distribution when considering a strong relevant feature~\citep{John1994,Yu2004}.
The sensor placement problem involves maximizing the area covered by a limited number of placed sensors.
The placement of the sensors becomes crucial because they are often operated for a long time after once installed.
We accordingly prefer to find a better solution as much as possible with a sufficient computation time.

Recently, \citet{Chen2015} proposed an $\textnormal{A}^\ast$~search algorithm \citep{Zeng2015} to obtain an optimal solution of the non-decreasing submodular function maximization problem.
Their algorithm computes an upper bound by a variant of variable fixing technique with $\mathrm{O}(n)$ oracle queries. 
\citet{Sakaue2018} improved the $\textnormal{A}^\ast$~search algorithm to obtain better upper bounds for the non-decreasing submodular function maximization with a knapsack constraint.
Their algorithms quickly find good upper bounds; however, the attained upper bounds are not often tight enough to prune nodes of the search tree effectively.
Therefore, their algorithms often process many nodes of the search tree until obtaining an optimal solution.

In this paper, we propose an efficient branch-and-bound algorithm for the non-decreasing submodular function maximization problem based on the following binary integer programming (BIP) formulation~\citep{Nemhauser1981}:
\begin{equation}
\label{eq:BIP}
\begin{array}{lll}
\textnormal{maximize} & z \\
\textnormal{subject to} & z  \leq  f(S) + \displaystyle\sum_{i \in N \setminus S} f(\{i\} \mid S)   \; y_{i}, \; S \in F, \\
& \displaystyle\sum_{i \in N } y_i \leq k,\\
& y_{i} \in \{0, 1\}, \; i \in N,
\end{array}
\end{equation}
where $f(T \mid S) = f(S \cup T) - f(S)$ for all $S, T \subseteq N$ and $F$ denotes the set of all feasible solutions $S \subseteq N$ satisfying the cardinality constraint $|S| \le k$.

\citet{Nemhauser1981} previously proposed an exact algorithm called the constraint generation algorithm based on the BIP formulation (\ref{eq:BIP}).
The size of the BIP formulation grows exponentially compared to $n$ since it has more than $\binom{n}{k}$ constraints.
To overcome this, they proposed the constraint generation algorithm that starts from a reduced BIP problem with a small subset of constraints taken from the constraints.
Their algorithm repeats solving a reduced BIP problem while adding a new constraint at each iteration.
Unfortunately, this is not efficient in practice because their algorithm requires to solve many reduced BIP problems.
They also proposed a branch-and-bound algorithm with solving linear programming (LP) relaxation problems of the reduced BIP problems to obtain upper bounds.
Their branch-and-bound algorithm however can not prune nodes of the search tree efficiently because the LP relaxation problems often give much worse upper bounds than the reduced BIP problems.

\citet{Kawahara2009} proposed an exact algorithm for updating a lower bound based on Tuy's cutting plane method~\citep{Tuy1964}. 
The algorithm reformulated the submodular function maximization problem (\ref{eq:subm}) by using the Lovasz extension \citep{Lovasz1983}.
It iteratively finds a feasible solution and a unique hyperplane to cut off a feasible subset which clearly does not include any better feasible solutions.
To obtain an upper bound, they introduced the constraint generation algorithm \citep{Nemhauser1981} adding the obtained feasible solution as a constraint.

Both of the exact algorithms~\citep{Nemhauser1981,Kawahara2009} often need to solve a large number of reduced BIP problems because of generating only one or two constraints at each iteration.
In this paper, to overcome this, we propose an improved constraint generation algorithm to add a promising set of constraints at each iteration.
We incorporate it into a branch-and-bound algorithm to attain good upper bounds while solving a smaller number of reduced BIP problems.
To improve the efficiency of the branch-and-bound algorithm, we also introduce a few techniques to attain good upper and lower bounds quickly.

\section{Existing algorithms}
We review the $\textnormal{A}^\ast$~search algorithms proposed by \citeauthor{Chen2015} (\citeyear{Chen2015}) and \citeauthor{Sakaue2018} (\citeyear{Sakaue2018}), and the constraint generation algorithm proposed by \citeauthor{Nemhauser1981} (\citeyear{Nemhauser1981}) for the non-decreasing submodular function maximization problem.

\subsection{$\textnormal{\textbf{A}}^{\ast}$~Search Algorithm}
We first define the search tree of the $\textnormal{A}^{\ast}$~search algorithm.
Each node $S$ of the search tree represents a feasible solution, where the root node is set to $\emptyset$.
The parent of a node $T$ is defined as $S = T \setminus \{ T_{\max} \}$, where $T_{\max}$ is an element $i \in T$ with the largest number.
For example, node $S = \{ 3 \}$ is the parent of node $T = \{ 3, 5 \}$, since $T \setminus \{ T_{\max} \}= \{ 3,5 \} \setminus \{ 5 \} = \{ 3 \} = S$. 

The $\textnormal{A}^{\ast}$~search algorithm employs a list $L$ to manage nodes of the search tree.
The value of a node $S$ is defined as $\bar{f}(S) = f(S) + h(S)$, where $h(\cdot)$ is a heuristic function.
We note that $\bar{f}(\cdot)$ give an upper bound of the optimal value of the problem~(\ref{eq:subm}) at the node $S$.

The initial feasible solution is obtained by the greedy algorithm \citep{Nemhauser1978,Minoux1978}.
The algorithm repeats to extract a node $S$ with the largest value $\bar{f}(\cdot)$ from the list $L$ and insert its children $T \in F$ into the list $L$ at each iteration.
Let $S \in F$ be a node extracted from the list $L$, and $S^{\ast}$ be the best feasible solution obtained so far.
The algorithm applies the greedy algorithm to the node $S$ for obtaining a feasible solution $S^{\prime} \in F$.
If $f(S^{\prime}) > f(S^{\ast})$ holds, then the algorithm replaces the best feasible solution $S^{\ast}$ with $S^{\prime}$.
Then, all children $T \in F$ of the node $S$ satisfying $\bar{f}(T) > f(S^{\ast})$ are inserted into the list $L$.
The algorithm repeats these procedures until the list $L$ becomes empty.

\begin{description}
\item[\underline{\textbf{Algorithm A$^{\ast}$}$(S)$}] 
\item[\textbf{Input:}] The initial feasible solution $S$.
\item[\textbf{Output:}] An optimal solution $S^{\ast}$.
\item[\textbf{Step~1:}] Set $L \leftarrow \{ \emptyset \}$ and $S^{\ast} \leftarrow S$.
\item[\textbf{Step~2:}] If $L = \emptyset$ holds, then output an optimal solution $S^{\ast}$ and exit. 
\item[\textbf{Step~3:}] Extract a node $S$ with the largest value $\bar{f}(\cdot)$ from the list $L$.
If $\bar{f}(\cdot) \leq f(S^{\ast})$ holds, then return to Step~2.
\item[\textbf{Step~4:}] Apply the greedy algorithm from the node $S$ to obtain $S^{\prime} \in F$.
If $f(S^{\prime}) > f(S^{\ast})$ holds, then set $S^{\ast} \leftarrow S^{\prime}$.
\item[\textbf{Step~5:}] Set $L \leftarrow L \cup \{ T \}$ for all children $T$ of the node $S$ satisfying $T \in F$ and $\bar{f}(T) > f(S^{\ast})$.
Return to Step~2.
\end{description}

We then illustrate two heuristic functions $h_{mod}$ \citep{Chen2015} and $h_{dom}$ \citep{Sakaue2018} applied to the $\textnormal{A}^{\ast}$~search algorithm.

\subsection{Upper bound with modular functions (MOD)}
\citet{Chen2015} proposed a heuristic function $h_{mod}$.
Let $S$ be the current node of the $\textnormal{A}^{\ast}$~search algorithm. 
We consider the following reduced problem of the problem~(\ref{eq:subm}) for obtaining $h(\cdot)$.
\begin{equation}
\label{eq:red}
\begin{array}{ll}
\textnormal{maximize} & f_{S}(T) \\
\textnormal{subject to} & T \subseteq N \setminus S^+, |T| \leq k - |S|,
\end{array}
\end{equation}
where $S^+ = \{i \in N \mid i \leq S_{\max} \}$ and $f_S(\cdot) = f(\cdot \mid S)$.
Let $T^\ast$ be an optimal solution of the reduced problem (\ref{eq:red}). 
By submodularity, we obtain $\sum_{i \in T} f_S(\{ i \}) \geq f_{S}(T)$ for any $T \subseteq N$ and the following inequality. 
\begin{equation}
\max_{T \subseteq N \setminus S^+, |T| \leq k - |S| } \displaystyle\sum_{i \in T} f_{S}(\{ i \}) \geq \sum_{i \in T^\ast} f_{S}( \{ i\} )  \geq  f_{S}(T^\ast).
\label{eq:VF}
\end{equation}
Since the reduced problem (\ref{eq:red}) is still NP-hard, we consider obtaining an upper bound of $f_{S}(T^\ast)$.
Let $\bar{S}^+$ be the non-increasing ordered set with respect to $f_{S}(\{ i \})$ for $i \in N \setminus S^+$.
We assume that $|S \cup \bar{S}^+| > k$, because we can obtain the upper bound by computing $f(S \cup \bar{S}^+)$ in otherwise.
Let $[p] = \{ 1,\dots, p\}$ and $\bar{S}^+_{[p]}$ denote the set of the first $p = k - |S|$ elements of the sorted set $\bar{S}^+$. 
We then define a heuristic function $h_{mod}$ by
\begin{equation}
h_{mod}(S) = \sum_{i \in \bar{S}^+_{[p]}} f_{S}(\{ i \}).
\end{equation}
If $f_S(\{ i \}) = 0$ holds for some $i \in \bar{S}^+_{[p]}$, then we conclude $f_{S}(\bar{S}^+_{[p]}) = f_{S}(T^\ast)$ by submodularity.
For a given node $S$, we compute an upper bound $\bar{f}(S) = f(S) + h_{mod}$.

\subsection{Upper bound with dominant elements (DOM)}
\citet{Sakaue2018} proposed another heuristic function $h_{dom}$.
We define $T$ as the ordered set of $p = k - |S|$ elements added to the current solution $S$ by the greedy algorithm.
Let $T_i$ and $T_{[i]}$ denote the $i$-th element of the ordered set $T$ and the set of the first $i$ elements of the ordered set $T$, respectively.
We define 
\begin{equation}
\beta_{[p]} =  
\left\{
\begin{array}{ll}
      0  & \mathrm{if} \; h_{mod}(S \cup T_{[i]}) = 0 \; \mathrm{for~some} \; i \in [p]\\
       \displaystyle\prod_{i=1}^p \beta_{i}  & \mathrm{otherwise},
    \end{array}
\right.
\end{equation}
where
\begin{equation}
    \beta_i = 1 - \displaystyle\frac{ f_{S}(T_{i} \mid T_{[i-1]} ) }{h_{mod}(S \cup T_{[i-1]})}, \quad i \in [p].
\end{equation}
We then define a heuristic function $h_{dom}$ by
\begin{equation}
h_{dom} = \displaystyle\frac{f_{S}(T_{[p]})}{(1- \beta_{[p]})}.
\end{equation}
The heuristic function holds $h_{dom} \geq f_{S}(T^\ast)$ for any given $S \subseteq N$. 
For a given node $S$, we compute an upper bound $\bar{f}(S) = f(S) + h_{dom}$.

\subsection{Constraint generation algorithm}
\citet{Nemhauser1981} have proposed an exact algorithm called the constraint generation algorithm starting from a reduced BIP problem with  a small subset of constraints taken from the constraints.
The algorithm repeats solving a reduced BIP problem while adding a new constraint at each iteration.
Given a set of feasible solutions $Q \subseteq F$, we define $\textnormal{BIP}(Q)$ as the following reduced BIP problem of the problem (\ref{eq:BIP}).
\begin{equation}
\label{eq:BIPQ}
\begin{array}{lll}
\textnormal{maximize} & z \\
\textnormal{subject to} & z  \leq  f(S) + \displaystyle\sum_{i \in N \setminus S} f(\{ i \} \mid S)  \; y_{i}, \; S \in Q,\\
& \displaystyle\sum_{i \in N} y_{i} \leq k,  \\
&  y_{i} \in \{0, 1\}, \; i \in N.
\end{array}
\end{equation}

The initial solution $S^{(0)}$ is obtained by the greedy algorithm \citep{Nemhauser1978,Minoux1978}.
Their algorithm starts with a set $Q=  \{S^{(0)} \}.$
We consider the algorithm at the $t$-th iteration.
The algorithm first solves $\textnormal{BIP}(Q)$ with $Q = \{S^{(0)},\dots,S^{(t-1)}\}$ to obtain an optimal solution $\boldsymbol{y}^{(t)} = (y_1^{(t)},\dots,y_n^{(t)})$ and the optimal value $z^{(t)}$ that gives an upper bound of the problem (\ref{eq:BIP}).
Let $S^{(t)}$ denote the optimal solution of $\textnormal{BIP}(Q)$ corresponding to $\boldsymbol{y}^{(t)}$, and $S^{\ast}$ denote the best feasible solution of the problem (\ref{eq:BIP}) obtained so far.
If $f(S^{(t)}) > f(S^{\ast})$ holds, then the algorithm replaces the best feasible solution  $S^{\ast}$ with $S^{(t)}$. 
If $z^{(t)} > f(S^{(t)})$ holds, the algorithm concludes $S^{(t)} \notin Q$ and add $S^{(t)}$ to $Q$, because $S^{(t)}$ does not satisfy any constraints of $\textnormal{BIP}(Q)$.
That is, the algorithm adds the following constraint to $\textnormal{BIP}(Q)$ for improving the upper bound $z^{(t)}$ of the optimal value of the problem (\ref{eq:BIP}).
\begin{equation}
z \le f(S^{(t)}) + \sum_{i \in N \setminus S^{(t)}} f(\{ i \} \mid S^{(t)}) \; y_i.
\end{equation}
The algorithm repeats these procedures until $z^{(t)}$ and $f(S^{\ast})$ meet.
We note that the value of $z^{(t)}$ is non-increasing along to the number of iterations and the algorithm must terminate after at most $\binom{n}{k}$ iterations.
\begin{description}
\item[\underline{Algorithm CG$(S^{(0)})$}]
\item[\textbf{Input:}] The initial feasible solution $S^{(0)}$. 
\item[\textbf{Output:}] An optimal solution $S^{\ast}$.
\item[\textbf{Step 1:}] Set $Q \leftarrow S ^{(0)}$, $S^{\ast} \leftarrow S^{(0)}$ and $t \leftarrow 1$. 
\item[\textbf{Step 2:}]  Solve $\textnormal{BIP}(Q)$.
Let $S^{(t)}$ and  $z^{(t)}$ be an optimal solution and the optimal value of $\textnormal{BIP}(Q)$, respectively.
\item[\textbf{Step 3:}]  If $ f(S^{(t)}) > f(S^{\ast})$ holds, then set $S^\ast  \leftarrow S^{(t)}$.
\item[\textbf{Step 4:}] If $z^{(t)} = f(S^{\ast})$ holds, then output an optimal solution $S^{\ast} $ and exit. Otherwise; (i.e., $z^{(t)} > f(S^{\ast}) \ge f(S^{(t)})$), set $Q \leftarrow Q \cup \{ S^{(t)}\}$, $t \leftarrow t +1$ and return to Step 2.
\end{description}

\section{Proposed algorithms}
The constraint generation algorithm \citep{Nemhauser1981} often solves a large number of reduced BIP problems because of generating only one constraint at each iteration. 
We accordingly propose an improved constraint generation algorithm to generate a promising set of constraints for attaining good upper bounds while solving a smaller number of reduced BIP problems.

\subsection{Improved constraint generation algorithm}
Let $\boldsymbol{y}^{(t)} = (y_1^{(t)},\dots,y_n^{(t)})$ and $z^{(t)}$ be an optimal solution and the optimal value of $\textnormal{BIP}(Q)$ at the $t$-th iteration of the constraint generation algorithm, respectively.
We note that $z^{(t)}$ gives an upper bound of the optimal value of the problem (\ref{eq:BIP}).
To improve the upper bound $z^{(t)}$, it is necessary to add a new feasible solution $S^{\prime} \in F$ to $Q$ satisfying the following inequality.
\begin{equation}
z^{(t)} > f(S^{\prime}) + \sum_{i \in N \setminus S^{\prime}} f(\{ i \} \mid S^{\prime}) \; y_i^{(t)}.
\end{equation}
For this purpose, we now consider the following problem to generate a new feasible solution $S^{\prime} \in F$ adding to $Q$ called the pricing problem.
\begin{equation}
\label{eq:pricing}
\begin{array}{ll}
\textnormal{minimize} &  f(S) + \displaystyle\sum_{i \in N \setminus S}f(\{i\} \mid S) \; y_i^{(t)} \\
\textnormal{subject to} & |S| \leq k, \; S \subseteq N. 
\end{array}
\end{equation}
If the optimal value of the pricing problem (\ref{eq:pricing}) is less than $z^{(t)}$, then we add an optimal solution $S^{\prime}$ of the pricing problem (\ref{eq:pricing}) to $Q$; otherwise, we conclude $z^{(t)}$ is the optimal value of the problem (\ref{eq:BIP}).
We repeat adding a new feasible solution $S^{\prime}$ obtained from the pricing problem (\ref{eq:pricing}) to $Q$ and solving the updated $\textnormal{BIP}(Q)$ until $z^{(t)}$ and $S^{\prime}$ meet.
This procedure is often called the column generation method \citep{Chvatal1983} which is used for LP problems with a huge number of constraints.
However, the computational cost to solve a pricing problem (\ref{eq:pricing}) is very expensive, almost the same as solving the problem (\ref{eq:subm}).
To overcome this, we propose an improved constraint generation algorithm to quickly generate a promising set of constraints. 

After solving $\textnormal{BIP}(Q)$, we obtain at least one feasible solution $S^{\natural} \in Q$ attaining the optimal value $z^{(t)}$ of $\textnormal{BIP}(Q)$ such that
\begin{equation}
\label{eq:tight}
z^{(t)} = f(S^{\natural}) + \displaystyle\sum_{i \in N \setminus S^{\natural}} f(\{i\} \mid S^{\natural}) \; y_i^{(t)}.
\end{equation}

Let $S^{(t)}$ be the optimal solution of of $\textnormal{BIP}(Q)$ corresponding to $\boldsymbol{y}^{(t)}$, where we assume $S^{(t)} \not\in Q$.
We then consider adding an element $j \in S^{(t)} \setminus S^{\natural}$ to $S^{\natural}$, and obtain the following inequality by submodularity:
\begin{equation}
\label{eq:add}
\begin{array}{lcl}
z^{(t)} & = &  f(S^{\natural}) + \displaystyle\sum_{i \in N \setminus S^{\natural}} f(\{i\} \mid S^{\natural}) \; y_i^{(t)}\\
& = & f(S^{\natural}) + f(\{ j \} \mid S^{\natural}) y_j^{(t)} + \displaystyle\sum_{i \in N \setminus (S^{\natural} \cup \{ j \}) }  f(\{i\} \mid S^{\natural}) y_i^{(t)}\\
& = & f(S^{\natural} \cup \{j\}) + \displaystyle\sum_{i \in N \setminus (S^{\natural} \cup \{j\} )} f(\{i\} \mid S^{\natural}) \; y_i^{(t)}\\
& \geq &  f(S^{\natural} \cup \{j\}) + \displaystyle\sum_{i \in N \setminus (S^{\natural} \cup \{j\} )} f( \{i\} \mid S^{\natural} \cup \{j\}) \; y_i^{(t)},
\end{array}
\end{equation}
where $y_j^{(t)} = 1$ due to $j \in S^{(t)}$.
From the inequality (\ref{eq:add}), we observe that it is preferable to add the element $j \in S^{(t)} \setminus S^{\natural}$ to $S^{\natural}$ for improving the upper bound $z^{(t)}$.
Here, we note that it is necessary to remove another element $i \in S^{\natural}$ if $|S^{\natural}| = k$ holds.

Based on this observation, we develop a heuristic algorithm to generate a set of new feasible solutions $S^{\prime} \in F$ for improving the upper bound $z^{(t)}$.
Given a set of feasible solutions $Q \subseteq F$, let $a_i$ be the number of feasible solutions $S \in Q$ including an element $i \in N$.
We define the occurrence rate $p_i$ of each element $i$ with respect to $Q$ as 
\begin{equation}
p_i = \frac{ a_i }{ \sum_{ j \in N}{ a_j}}.
\end{equation}
For each element $i \in S^{\natural} \cup S^{(t)}$, we set a random value $r_i$ satisfying $0 \leq r_i \leq p_i$.
If there are multiple feasible solutions $S^{\natural} \in Q$ satisfying the equation (\ref{eq:tight}), then we select one of them at random.
We take the $k$ largest elements $i \in S^{\natural} \cup S^{(t)}$ with respect to the value $r_i$ to generate a feasible solution $S^{\prime} \in F$.
\begin{description}
\item[\underline{Algorithm SUB-ICG$(Q, S^{(t)}, \lambda)$}]
\item[Input:] A set of feasible solutions $Q \subseteq F$.
A feasible solution $S^{(t)} \not\in Q$.
The number of feasible solutions to be generated $\lambda$.
\item[Output:] A set of feasible solutions $Q^{\prime} \subseteq F$.
\item[Step~1:] Set $Q^{\prime} \leftarrow \emptyset$ and $h \leftarrow 1$.
\item[Step~2:] Select a feasible solution $S^{\natural} \in Q$ satisfying the equation (\ref{eq:tight}) at random.
Set a random value $r_i$ $(0 \le r_i \le p_i)$ for $i \in S^{\natural} \cup S^{(t)}$.
\item[Step~3:] Take the $k$ largest elements $i \in S^{\natural} \cup S^{(t)}$ with respect to $r_i$ to generate a feasible solution $S^{\prime} \in F$.
\item[Step~4:] If $S^{\prime} \not\in Q^{\prime}$ holds, then set $Q^{\prime} \leftarrow Q^{\prime} \cup \{ S^{\prime} \}$ and $h \leftarrow h + 1$.
\item[Step~5:] If $h = \lambda$ holds, then output $Q^{\prime}$ and exit.
Otherwise, return to Step~2.
\end{description}

We summarize the improved constraint generation algorithm as follows, in which we define $Q$ as the set of feasible solutions obtained by solving $\textnormal{BIP}(Q)$ and $Q^+$ as the set of feasible solutions generated by $\textnormal{SUB-ICG}(Q,S^{(t)}, \lambda)$.

\begin{description}
\item[\underline{Algorithm ICG$(S^{(0)},\lambda)$}]
\item[Input:] The initial feasible solution $S^{(0)}$.
The number of feasible solutions to be generated at each iteration $\lambda$.
\item[Output:] An optimal solution $S^{\ast}$.
\item[Step~1:] Set $Q \leftarrow \{ S^{(0)}$\}, $Q^{+} \leftarrow \{ S^{(0)}$\}, $S^{\ast} \leftarrow S^{(0)}$ and $t \leftarrow 1$.
\item[Step~2:] Solve $\textnormal{BIP}(Q^{+})$.
Let $S^{(t)}$ and $z^{(t)}$ be an optimal solution and the optimal value of $\textnormal{BIP}(Q^{+})$, respectively.
\item[Step~3:] If $f(S^{(t)}) > f(S^{\ast})$ holds, then set $S^{\ast} \leftarrow S^{(t)}$.
\item[Step~4:] If $z^{(t)} = f(S^{\ast})$ holds, then output an optimal solution $S^{\ast}$ and exit.
\item[Step~5:] Set $Q \leftarrow Q \cup \{ S^{(t)} \}$, $Q^{+} \leftarrow Q^{+} \cup \{ S^{(t)} \} \cup \textnormal{SUB-ICG}(Q,S^{(t)}, \lambda)$ and  $t \leftarrow t + 1$. 
\item[Step~6:] For each feasible solution $S \in \textnormal{SUB-ICG}(Q,S^{(t)},\lambda)$, if $f(S) > f(S^{\ast})$ holds, then set $S^{\ast} \leftarrow S$, and return to Step~2.
\end{description}

\subsection{Branch-and-bound algorithm}
We propose a branch-and-bound algorithm incorporating the improved constraint generation algorithm.
Branch-and-bound algorithm is an exact algorithm for optimization problems, which consists of an implicit enumeration using a search tree and a pruning mechanism using relaxation problems.

We first define the search tree of the branch-and-bound algorithm. 
Each node $(S^0, S^1)$ of the search tree consists a pair of sets $S^0$ and $S^1$, where elements $i \in S^0$ (resp., $i \in S^1$) correspond to variables fixed to $y_i = 0$ (resp., $y_i = 1$) of the problem (\ref{eq:BIP}). 
The root node is set to $(\emptyset, \emptyset)$.
Each node $(S^0,S^1)$ has two children $(S^0 \cup \{ i^{\ast} \},S^1)$ and $(S^0, S^1 \cup \{ i^{\ast} \})$, where $i^{\ast} = \mathrm{argmax}_{i \in N \setminus (S^0 \cup S^1)} f(S^1 \cup \{ i \})$.

The branch-and-bound algorithm employs a stack list $L$ to manage nodes of the search tree.
The value of a node $(S^0,S^1)$ is defined as the optimal value $z^{(S^0,S^1)}$ of the following reduced BIP problem $\textnormal{BIP}(Q^+, S^0, S^1)$:
\begin{equation}
\begin{array}{lll}
\textnormal{maximize} & z\\
\textnormal{subject to} & \multicolumn{2}{l}{z \le f(S) + \displaystyle\sum_{i \in N \setminus S} f( \{ i \} \mid S)\; y_i, \; S \in Q^+,}\\
& \multicolumn{2}{l}{\displaystyle\sum_{i \in N \setminus (S^0 \cup S^1)} y_i \le k - |S^1|,}\\
& y_i \in \{ 0, 1 \}, & i \in N \setminus (S^0 \cup S^1),\\
& y_i = 0, & i \in S^0,\\
& y_i = 1, & i \in S^1,
\end{array}
\end{equation}
where $Q^+$ is the set of feasible solution generated by the improved constraint generation algorithm so far.
We note that $z^{(S^0,S^1)}$ gives an upper bound of the optimal value of the problem (\ref{eq:BIP}) at the node $(S^0,S^1)$; i.e., under the condition that $y_i = 0$ $(i \in S^0)$ and $y_i = 1$ $(i \in S^1)$.

We start with a pair of sets $Q = \{ S \}$ and $Q^+ = \{ S \}$, where $S$ is the initial feasible solutions obtained by the greedy algorithm \citep{Nemhauser1981,Minoux1978}.
To obtain good upper and lower bounds quickly, we first apply the first $k$ iterations of the improved constraint generation algorithm.
We then repeat to extract a node $(S^0,S^1)$ from the top of the stack list $L$ and insert its children into the top of the stack list $L$ at each iteration.

Let $(S^0,S^1)$ be a node extracted from the stack list $L$, and $S^{\ast}$ be the best feasible solution of the problem (\ref{eq:BIP}) obtained so far.
We first solve $\textnormal{BIP}(Q^+, S^0, S^1)$ to obtain an optimal solution $S^{(S^0,S^1)}$ and the optimal value $z^{(S^0,S^1)}$.
We then generate a set of feasible solution $\textnormal{SUB-ICG}(Q,S^{(S^0,S^1)},\lambda)$ by the heuristic algorithm.
For each feasible solution $S^{\prime} \in \{ S^{(S^0,S^1)} \} \cup \textnormal{SUB-ICG}(Q,S^{(S^0,S^1)},\lambda)$, if $f(S^{\prime}) > f(S^{\ast})$ holds, then we replace the best feasible solution $S^{\ast}$ with $S^{\prime}$.
If $z^{(S^0,S^1)} > f(S^{\ast})$ holds, then we insert the two children $(S^0 \cup \{ i^{\ast} \}, S^1)$ and $(S^0, S^1 \cup \{ i^{\ast} \})$ into the top of the stack list $L$ in this order.

To decrease the number of reduced BIP problems to be solved in the branch-and-bound algorithm, we keep the optimal value $z^{(S^0,S^1)}$ of $\textnormal{BIP}(Q^+,S^0,S^1)$ as an upper bound $\bar{z}^{(S^0 \cup \{ i^{\ast} \}, S^1)}$ (resp., $\bar{z}^{(S^0,S^1 \cup \{ i^{\ast} \})}$) of the child $(S^0 \cup \{ i^{\ast} \}, S^1)$ (resp., $(S^0, S^1 \cup \{ i^{\ast} \})$) when inserted to the stack list $L$.
If $\bar{z}^{(S^0,S^1)} \le f(S^{\ast})$ holds when we extract a node $(S^0,S^1)$ from the stack list $L$, then we can prune the node $(S^0,S^1)$ without solving $\textnormal{BIP}(Q^+,S^0,S^1)$.
We set the upper bound $\bar{z}^{(\emptyset,\emptyset)}$ of the root node $(\emptyset,\emptyset)$ to $\infty$. 
We repeat these procedures until the stack list $L$ becomes empty.

\begin{description}
\item[\underline{Algorithm BB-ICG$(S,\lambda)$}]
\item[\textbf{Input:}] The initial feasible solution $S$.
The number of feasible solutions to be generated at each node $\lambda$. 
\item[\textbf{Output:}] An optimal solution $S^{\ast}$.
\item[Step~1:] Set $L \leftarrow \{ (\emptyset,\emptyset) \}$, $\bar{z}^{(\emptyset,\emptyset)} \leftarrow \infty$, $Q \leftarrow \{ S\}$,  $Q^+ \leftarrow \{ S \}$ and $S^{\ast} \leftarrow S$.
\item[Step~2:] Apply the first $k$ iterations of $\textnormal{ICG}(S,\lambda)$ to update the sets $Q$ and $Q^+$ and the best feasible solution $S^{\ast}$.
\item[Step~3:] If $L = \emptyset$ holds, then output an optimal solution $S^{\ast}$ and exit.
\item[Step~4:] Extract a node $(S^0,S^1)$ from the top of the stack list $L$.
If $\bar{z}^{(S^0,S^1)} \le f(S^{\ast})$ holds, then return to Step~3.
\item[Step~5:] Solve $\textnormal{BIP}(Q^+,S^0,S^1)$.
Let $S^{(S^0,S^1)}$ and $z^{(S^0,S^1)}$ be an optimal solution and the optimal value of $\textnormal{BIP}(Q^+,S^0,S^1)$, respectively.
\item[Step~6:] Set $Q \leftarrow Q \cup \{ S^{(S^0,S^1)} \}$, $Q^+ \leftarrow Q^+ \cup \{ S^{(S^0,S^1)} \} \cup \textnormal{SUB-ICG}(Q,S^{(S^0,S^1)},\lambda)$.
\item[Step~7:] For each feasible solution $S^{\prime} \in \{ S^{(S^0,S^1)} \} \cup \textnormal{SUB-ICG}(Q,S^{(S^0,S^1)},\lambda)$, if $f(S^{\prime}) > f(S^{\ast})$ holds, then set $ S^{\ast} \leftarrow  S$.
\item[Step~8:] If $ z^{(S^0,S^1)} \leq f(S^{\ast})$, then return to Step 3.
\item[Step~9:] Set $L \leftarrow L \cup \{ (S^0 \cup \{ i^{\ast} \}, S^1), (S^0, S^1 \cup \{ i^{\ast} \}) \}$ and $\bar{z}^{(S^0 \cup \{ i^{\ast} \}, S^1)} \leftarrow z^{(S^0,S^1)}$ and $\bar{z}^{(S^0, S^1\cup \{ i^{\ast} \})} \leftarrow z^{(S^0,S^1)}$, where $i^{\ast} = \mathrm{argmax}_{i \in N \setminus (S^0 \cup S^1)} f(S^1 \cup \{ i \})$ and $|S^1| \le k-1$ hold.
Return to Step~3.
\end{description}

\subsection{Improved branch-and-bound algorithm}
We finally propose an improved branch-and-bound algorithm that introduces (i)~the heuristic function $h_{dom}$ to obtain a good upper bound quickly and (ii)~a local search algorithm to improve the lower bound from the best feasible solution obtained so far. 

The computational cost of evaluating the heuristic function $h_{dom}$ is much cheaper than that of solving a reduced BIP problem.
We accordingly compute an upper bound $\bar{f}(S^1) = f(S^1) + h_{dom}(S^1)$ of the problem (\ref{eq:BIP}) at the node $(S^0,S^1)$ before solving $\textnormal{BIP}(Q^+,S^0,S^1)$, where we set $S^{1+} = S^1 \cup S^0$ for computing $h_{dom}(S^1)$.

To improve the efficiency of the branch-and-bound algorithm, it is also important to improve the lower bound from the best feasible solution $S^{\ast}$ obtained so far.
We accordingly introduce a simple local search at each node $(S^0,S^1)$ of the branch-and-bound algorithm.
We first apply the greedy algorithm from $S^1$ to obtain an initial feasible solution $S \in F$, where we only consider to add an element $i \in N \setminus (S^0 \cup S^1)$ at each iteration. 
We then repeatedly replaces $S$ with a better feasible solution $S^{\prime}$ in its neighborhood $\mathrm{NB}(S)$ until no better feasible solution
is found in $\mathrm{NB}(S)$. 
For a given feasible solution $S \in F$, we define an exchange neighborhood as 
$\mathrm{NB}(S) = \{ S^{\prime} \subseteq N \mid S \setminus \{ i \} \cup \{ j \}, i \in S \setminus S^1, j \in N \setminus (S \cup S^0) \}$.

\begin{description}
\item[\underline{Algorithm LS$(S^0,S^1)$}]
\item[Input:] A node of the branch-and-bound algorithm $(S^0,S^1)$. 
\item[Output:] A feasible solution $S$. 
\item[Step~1:] Apply the greedy algorithm from $S^1$ to obtain an initial feasible solution $S$.
\item[Step~2:] Find the best feasible solution $S^{\prime} \in \mathrm{NB}(S)$.
If $f(S^{\prime}) > f(S)$ holds, then set $S \leftarrow S^{\prime}$ and return to Step~2; otherwise, output $S$ and exit.
\end{description}

The improved branch-and-bound algorithm is described by replacing Steps~4 and 8 of the branch-and-bound algorithm as follows.
\begin{description}
\item[Step~4$^{\prime}$:] Extract a node $(S^0,S^1)$ from the top of the list $L$.
Set $S \leftarrow \textnormal{LS}(S^0,S^1)$.
If $f(S) > f(S^{\ast})$ holds, then set $S^{\ast} \leftarrow S$.
If $\min \{ \bar{f}(S^1), \bar{z}^{(S^0,S^1)} \} \le f(S^{\ast})$ holds, then return to Step~3.
\item[Step~8$^\prime$:] If $ \min \{ \bar{f}(S^1), z^{(S^0 , S^1)} \} \leq f(S^{\ast})$, then return to Step 3.
\end{description}

\section{Computational Results}
We tested three existing algorithms and three proposed algorithms: (i)~the $\textnormal{A}^{\ast}$~search algorithm with the heuristic function $h_{mod}$ ($\textnormal{A}^{\ast}$-MOD), (ii)~the $\textnormal{A}^{\ast}$~search algorithm with the heuristic function $h_{dom}$ ($\textnormal{A}^{\ast}$-DOM), (iii)~the constraint generation algorithm (CG), (iv)~the improved constraint generation algorithm (ICG), (v)~the branch-and-bound algorithm (BB-ICG) and (vi)~the improved branch-and-bound algorithm (BB-ICG+).
All algorithms were tested on a personal computer with a 4.0~GHz Intel Core i7 processor and 32~GB memory.
For ICG, BB-ICG and BB-ICG+, we use a mixed integer programming (MIP) solver called CPLEX~12.6 for solving BIP problems, and the number of feasible solutions to be generated at each iteration $\lambda$ is set to $10k$ based on computational results of preliminary experiments.

We report computational results for two types of well-known benchmark instances called \textit{facility location} (LOC) and \emph{weighted coverage} (COV)~\citep{Kawahara2009,Sakaue2018}.

\noindent\textbf{Facility location (LOC)} 
We are given a set of $n$ locations $N = \{ 1,\dots,n \}$ and a set of $m$ clients $M = \{ 1,\dots,m \}$.
We consider to select a set of $k$ locations to build facilities.
We define $g_{ij} \geq 0$ as the benefit of $i$-th client attaining from the facility at $j \in N$.
We select a set of locations $S \subseteq N$ to built the facilities.
Each client $i \in M$ attains the benefit from the most beneficial facility.
The total benefit for the clients is defined as 
\begin{equation}
f(S) = \displaystyle\sum_{i \in M} \max_{j \in S} g_{ij}.
\end{equation}

\noindent\textbf{Weighted coverage (COV)}
We are given a set of $m$ items $M = \{ 1,\dots,m \}$ and a set of $n$ sensors $N = \{ 1,\dots,n \}$.
Let $M_j \subseteq M$ be the subset of items covered by an item $j \in N$, and $w_i \ge 0$ be a weight of item $i \in M$.
The total weighted coverage for the items is defined as
\begin{equation}
f(S) = \sum_{i \in \bigcup_{j \in S} M_j} w_i.
\end{equation}

We tested all algorithms for 24 classes of randomly generated instances that are characterized by several parameters.
We set $n=30,\dots,55$, $m = n + 1$ and $k= 5,8$ for both types of instances.
For the facility location instances, $g_{ij}$ is a random value taken from interval $[0,1]$.
For the weighted coverage instances, each sensor $j \in N$ randomly covers each item $i \in M$ with probability $0.07$ and $w_i$ is a random value taken from interval $[0,1]$.
For each class, five instances were generated and tested.
For all instances, we set the time limit to one hour (3600 seconds).

\begin{table*}[!tb]
\caption{Computation time (in seconds) of the proposed algorithms and the existing algorithms.}
\centering
{\scriptsize
\begin{tabular}{@{}c@{~}r@{~}r@{\,~~}r@{\,~~}r@{\,~~}r@{\,~~}r@{\,~~}r@{\,~~}r@{}}
\\
 \hline
\multicolumn{1}{c}{Type} & \multicolumn{1}{c}{$n$} & \multicolumn{1}{c}{$k$} & \multicolumn{1}{c}{A$^{\ast}$-MOD} & \multicolumn{1}{c}{A$^{\ast}$-DOM} & \multicolumn{1}{c}{CG} & \multicolumn{1}{c}{ICG} & \multicolumn{1}{c}{BB-ICG} & \multicolumn{1}{c}{BB-ICG+}       \\
 \hline
 & $30$ & $5$  &  $1.60 \times 10^1$ $(5)$ &  $3.22 \times 10^1$ $(5)$  &  $ 8.85 \times 10^0$ $(5)$ & $\boldsymbol{3.61 \times 10^0}$ $(5)$ &  $4.96 \times 10^0$  $(5)$ & $5.57 \times 10^0$ $(5)$  \\
 &  $35$ & $5$  &  $3.88 \times 10^1$ $(5)$  &  $7.59 \times 10^1$ $(5)$ & $2.85 \times 10^1$ $(5)$ & $\boldsymbol{1.09 \times 10^1}$ $(5)$ & $1.56 \times 10^1$ $(5)$ & $1.54 \times 10^1$ $(5)$   \\
  &  $40$ & $5$ & $8.27 \times 10^1$ $(5)$ & $1.53 \times 10^2$ $(5)$ & $4.05 \times 10^2$ $(5)$ & $6.24 \times 10^1$ $(5)$ & $5.38 \times 10^1$ $(5)$ &  $\boldsymbol{5.01 \times 10^1}$ $(5)$   \\
  LOC &  $45$ & $5$  &  $1.72 \times 10^2$ $(5)$  &  $2.97 \times 10^2$ $(5)$  & $1.63 \times 10^3$ $(5)$ & $4.38 \times 10^2$ $(5)$ & $1.39 \times 10^2$ $(5)$ & $\boldsymbol{1.31 \times 10^2}$ $(5)$  \\
    &  $50$ & $5$  &  $3.24 \times 10^2$ $(5)$  &  $5.25 \times 10^2$ $(5)$ & $>$ $1.64 \times 10^3$ $(3)$ & $6.33 \times 10^2$ $(5)$ & $2.47 \times 10^2$ $(5)$ & $\boldsymbol{1.81 \times 10^2}$ $(5)$  \\
   &  $55$ & $5$  &  $5.55 \times 10^2$ $(5)$   &  $8.84 \times 10^2$ $(5)$ & $>$ $2.69 \times 10^3$ $(2)$ & $>$ $1.36 \times 10^3$ $(4)$ & $3.99 \times 10^2$ $(5)$ & $\boldsymbol{3.86 \times 10^2}$ $(5)$   \\
    \hline
     &  $30$ & $8$  &  $5.11 \times 10^2$ $(5)$ & $9.37 \times 10^2$ $(5)$  & $9.28 \times 10^0$ $(5)$  & $\boldsymbol{1.13 \times 10^0}$ $(5)$ &  $5.88 \times 10^0$ $(5)$ & $5.45 \times 10^0$ $(5)$   \\
     
      &  $35$ & $8$  &  $2.12 \times 10^3$ $(5)$  & $>$ $3.30 \times 10^3$ $(2)$ & $2.02 \times 10^1$ $(5)$  & $\boldsymbol{6.70 \times 10^0}$ $(5)$  & $1.07 \times 10^1$ $(5)$ & $9.60 \times 10^0$  $(5)$  \\
      
      &  $40$ & $8$  & $>$ $3.60 \times 10^3$ $(0)$  & $>$ $3.60 \times 10^3$ $(0)$  & $>$ $1.10 \times 10^3$ $(4)$ & $5.32 \times 10^2$ $(5)$ & $1.56 \times 10^2$ $(5)$ & $\boldsymbol{1.36 \times 10^2 }$ $(5)$   \\
      
   LOC    &  $45$ & $8$  & $>$ $3.60 \times 10^3$ $(0)$   & $>$ $3.60 \times 10^3$ $(0)$   & $>$ $2.90 \times 10^3$ $(1)$ & $>$ $1.23 \times 10^3$ $(4)$  & $7.97 \times 10^2$ $(5)$ & $\boldsymbol{6.76 \times 10^2}$ $(5)$   \\
   
        &  $50$ & $8$  & $>$ $3.60 \times 10^3$ $(0)$    & $>$ $3.60 \times 10^3$  $(0)$ & $>$ $3.60 \times 10^3$  $(0)$ & $>$ $ 2.22 \times 10^3$ $(3)$  & $1.10 \times 10^3$ $(5)$ & $\boldsymbol{1.01 \times 10^3}$ $(5)$   \\
        
         &  $55$ & $8$  & $>$ $3.60 \times 10^3$ $(0)$  & $>$ $3.60 \times 10^3$  $(0)$  & $>$ $3.60 \times 10^3$  $(0)$ &$>$ $3.60 \times 10^3$ $(0)$ & $>$ $2.24 \times 10^3$ $(3)$ & $\boldsymbol{1.88 \times 10^3}$ $(5)$   \\
       \hline
    &  $30$ & $5$  &  $4.67 \times 10^{-1}$ $(5)$   &  $1.18 \times 10^0$ $(5)$  & $1.06 \times 10^0$ $(5)$ & $5.97 \times 10^{-1}$ $(5)$ &  $5.91 \times 10^{-1}$  $(5)$  & $\boldsymbol{4.43 \times 10^{-1}}$  $(5)$  \\ 
    
     &  $35$ & $5$  &  $1.02 \times 10^0$ $(5)$  &  $2.84 \times 10^0$ $(5)$  & $1.56 \times 10^0$ $(5)$ & $9.34 \times 10^{-1}$ $(5)$ &  $1.05 \times 10^{0}$ $(5)$  & $\boldsymbol{7.73 \times 10^{-1}}$  $(5)$  \\ 
     
      &  $40$ & $5$  &  $1.58 \times 10^0$ $(5)$ &  $5.41 \times 10^0$ $(5)$  & $1.00 \times 10^1$ $(5)$  & $2.58 \times 10^0$ $(5)$  &  $3.86 \times 10^0 $ $(5)$  & $\boldsymbol{1.91 \times 10^0}$  $(5)$   \\ 
      
  COV   &  $45$ & $5$  &  $2.37 \times 10^0$ $(5)$  & $8.36 \times 10^0$ $(5)$ & $1.45 \times 10^1$ $(5)$& $3.39 \times 10^0$ $(5)$ &   $4.51 \times 10^0 $  $(5)$ & $\boldsymbol{3.34 \times 10^0}$  $(5)$  \\ 
  
        &  $50$ & $5$  & $4.93 \times 10^0$ $(5)$  & $1.72 \times 10^1$ $(5)$ & $3.22 \times 10^1$ $(5)$ & $6.35 \times 10^0$ $(5)$ & 
        $7.56 \times 10^0 $ $(5)$ & $\boldsymbol{5.03 \times 10^0}$  $(5)$  \\ 
        
      &  $55$ & $5$  &  $7.40 \times 10^0$ $(5)$  &  $2.71 \times 10^1$ $(5)$ & $2.24 \times 10^1$ $(5)$ & $5.49 \times 10^0$ $(5)$ & $7.02 \times 10^0$  $(5)$ & $\boldsymbol{4.85 \times 10^0}$  $(5)$   \\ 
         \hline
         
     & $30$ & $8$  &  $1.09 \times 10^1$ $(5)$ & $2.41 \times 10^1$ $(5)$ & $8.65 \times 10^{-1}$ $(5)$ & $\boldsymbol{7.50 \times 10^{-1}}$ $(5)$ 
     & $9.34 \times 10^{-1}$ $(5)$ & $8.62 \times 10^{-1}$ $(5)$ \\
     
          & $35$ & $8$  &  $3.45 \times 10^1$ $(5)$  &  $6.79 \times 10^1$ $(5)$ & $2.04 \times 10^{0}$ $(5)$ & $\boldsymbol{1.28 \times 10^{0}}$ $(5)$ 
          & $1.41 \times 10^{0}$ $(5)$ &  $1.37 \times 10^{0}$ $(5)$  \\  
          
           & $40$ & $8$  &  $3.32 \times 10^1$ $(5)$  &  $1.49 \times 10^2$ $(5)$ & $4.52 \times 10^{0}$ $(5)$ & $2.40 \times 10^{0}$ $(5)$ & $\boldsymbol{2.12 \times 10^{0}}$ $(5)$ & $2.75 \times 10^{0}$ $(5)$  \\  
           
    COV    & $45$ & $8$  &  $1.05 \times 10^2$ $(5)$ & $5.20 \times 10^2$ $(5)$ &  $2.33 \times 10^1$ $(5)$ & $4.69 \times 10^0 $ $(5)$  & 
    $\boldsymbol{4.64 \times 10^0 }$ $(5)$  & $5.03 \times 10^0$  $(5)$ \\ 
    
         & $50$ & $8$  & $2.72 \times 10^2$  $(5)$  & $>$ $1.11 \times 10^3$  $(4)$ & $7.26 \times 10^1$ $(5)$ & $\boldsymbol{5.94 \times 10^0}$ $(5)$ & $6.32 \times 10^0 $ $(5)$  & $6.84 \times 10^0$ $(5)$  \\ 
         
        & $55$ & $8$  & $3.97 \times 10^2$ $(5)$  & $>$ $1.65 \times 10^3$  $(4)$ & $1.77 \times 10^2$ $(5)$ & $\boldsymbol{7.83 \times 10^0}$ $(5)$   & $9.04 \times 10^0$ $(5)$ & $9.21 \times 10^0$ $(5)$  \\ 
             \hline
\multicolumn{3}{l}{Avg. factor} & $>21.81$ $(100)$ & $>48.81$ $(95)$ & $>5.22$ $(100)$ & $>1.51$ $(111)$ & $>1.17$ $(118)$ & $1.00$ $(120)$ \\
\hline
 \end{tabular}
}
 \label{tab:time}
\end{table*}

\begin{table*}[!tb]
\caption{Number of processed nodes by the proposed algorithms and the existing algorithms.}
\centering
{\scriptsize
\begin{tabular}{lrrrrrrrr}
\\
\hline
\multicolumn{1}{c}{Type} & \multicolumn{1}{c}{$n$} & \multicolumn{1}{c}{$k$} & \multicolumn{1}{c}{A$^{\ast}$-MOD} & \multicolumn{1}{c}{A$^{\ast}$-DOM} & \multicolumn{1}{c}{BB-ICG} & \multicolumn{1}{c}{BB-ICG+}    \\
 \hline
 & $30$ & $5$  & $1.78 \times 10^4$ $(5)$  &  $1.45 \times 10^4$ $(5)$ & $2.70 \times 10^1$ $(5)$ &  $\boldsymbol{2.62 \times 10^1}$ $(5)$ \\
 &  $35$ & $5$  & $3.47 \times 10^4$ $(5)$ & $2.92 \times 10^4$ $(5)$  & $6.70 \times 10^1$ $(5)$ & $\boldsymbol{5.42 \times 10^1}$ $(5)$\\
 & $40$ & $5$  & $5.74 \times 10^4$ $(5)$  & $4.78 \times 10^4$ $(5)$ & $1.24 \times 10^2$ $(5)$ & $\boldsymbol{1.01 \times 10^2}$ $(5)$ \\
 LOC &  $45$ & $5$  &  $9.89 \times 10^4$ $(5)$ & $8.25 \times 10^4$ $(5)$ & $1.71 \times 10^2$ $(5)$ & $\boldsymbol{1.52 \times 10^2}$ $(5)$ \\
  &  $50$ & $5$  &  $1.53 \times 10^5$ $(5)$   & $1.26 \times 10^5$ $(5)$ & $2.37 \times 10^2$ $(5)$ & $\boldsymbol{1.73 \times 10^2}$ $(5)$ \\
 &  $55$ & $5$  & $2.13 \times 10^5$ $(5)$  & $1.82 \times 10^5$ $(5)$ & $2.86 \times 10^2$ $(5)$ & $\boldsymbol{2.68 \times 10^2}$ $(5)$ \\
   \hline
  &  $30$ & $8$  & $4.95 \times 10^5$ $(5)$   & $2.05 \times 10^5$ $(5)$ & $1.02 \times 10^1 $ $(5)$ & $\boldsymbol{7.40 \times 10^0}$ $(5)$ \\
  
  &  $35$ & $8$  &     $1.53 \times 10^6$ $(5)$  & $>$ $5.40 \times 10^5$ $(2)$ & $1.22 \times 10^1$ $(5)$ &  $\boldsymbol{9.40 \times 10^0}$ $(5)$\\
  
  &  $40$ & $8$  & $>$ $1.88 \times 10^6$  $(0)$ &  $>$ $3.56 \times 10^5$   $(0)$ & $7.10 \times 10^1$ $(5)$ & $\boldsymbol{6.50 \times 10^1}$ $(5)$      \\
  
 LOC &  $45$ & $8$  & $>$ $1.50 \times 10^6$   $(0)$ & $>$ $ 2.47 \times 10^5 $   $(0)$ & $1.66 \times 10^2$ $(5)$ & $\boldsymbol{1.30 \times 10^2}$ $(5)$ \\
 
 & $50$ & $8$ & $>$ $1.28 \times 10^6$  $(0)$ & $>$ $1.79 \times 10^5$   $(0)$ & $2.15 \times 10^2 $ $(5)$& $\boldsymbol{1.68 \times 10^2}$ $(5)$ \\
 
  &  $55$ & $8$ & $>$ $1.12 \times 10^6$   $(0)$ & $>$ $1.36 \times 10^5$  $(0)$ & $>$ $2.81 \times 10^2 $ $(3)$ & $\boldsymbol{1.80 \times 10^2}$ $(5)$ \\
   \hline
 & $30$ & $5$ & $8.59 \times 10^2$ $(5)$  &  $6.66 \times 10^2$ $(5)$ & $3.40 \times 10^0$ $(5)$ & $\boldsymbol{2.60 \times 10^0}$ $(5)$\\
 
 & $35$ & $5$  &   $1.65 \times 10^3$ $(5)$  &  $1.39 \times 10^3$ $(5)$ & $8.60 \times 10^0$ $(5)$ & $\boldsymbol{3.80 \times 10^0}$ $(5)$  \\
 
 &  $40$ & $5$  &   $2.05 \times 10^3$ $(5)$ &  $2.02 \times 10^3$ $(5)$ & $3.34 \times 10^1$ $(5)$ & $\boldsymbol{1.34 \times 10^1}$  $(5)$\\
 
 COV &  $45$ & $5$  &  $ 2.47 \times 10^3$ $(5)$  &  $2.44 \times 10^3$ $(5)$ & $2.66 \times 10^1$ $(5)$ & $\boldsymbol{1.74 \times 10^1}$ $(5)$ \\
 
  &  $50$ & $5$  &  $4.42 \times 10^3$ $(5)$   &  $4.41 \times 10^3$ $(5)$ & $3.70 \times 10^1$ $(5)$ & $\boldsymbol{2.34 \times 10^1}$  $(5)$     \\
  
 &  $55$ & $5$  & $5.52 \times 10^3$ $(5)$ & $5.39 \times 10^3$ $(5)$ & $2.62 \times 10^1$ $(5)$ & $\boldsymbol{1.74 \times 10^1}$ $(5)$\\
   \hline
  &  $30$ & $8$  & $1.70 \times 10^4$ $(5)$ &  $7.76 \times 10^3$ $(5)$ & $\boldsymbol{2.60 \times 10^0}$ $(5)$ & $3.00 \times 10^0$ $(5)$ \\
  
  &  $35$ & $8$ & $4.62 \times 10^4$ $(5)$ & $1.65 \times 10^4$ $(5)$ & $3.80 \times 10^0$ $(5)$ & $\boldsymbol{2.60 \times 10^0}$ $(5)$ \\
  
  &  $40$ & $8$ & $3.63 \times 10^4$ $(4)$  &  $2.72 \times 10^4$ $(5)$ & $\boldsymbol{2.20 \times 10^0}$ $(5)$ & $3.40 \times 10^0$ $(5)$  \\
  
COV  & $45$ & $8$ & $9.30 \times 10^4$ $(5)$  & $8.33 \times 10^4$   $(5)$ & $\boldsymbol{4.60 \times 10^0}$ $(5)$ & $5.80 \times 10^0$ $(5)$ \\

 & $50$ & $8$ & $2.02 \times 10^5$ $(5)$ &  $>$ $1.40 \times 10^5$  $(4)$ & $\boldsymbol{3.00 \times 10^0}$ $(5)$ & $3.40 \times 10^0$ $(5)$ \\
 
  &  $55$ & $8$ & $ 2.47 \times 10^5$  $(5)$   & $>$ $ 1.53 \times 10^5$  $(4)$ & $\boldsymbol{4.60 \times 10^0}$ $(5)$ & $5.00 \times 10^0$ $(5)$ \\
   \hline
   \multicolumn{3}{l}{Avg. factor} & $>18695.95$ $(100)$ & $>8431.07$ $(95)$ & $>1.30$ $(118)$ & $1.00$ $(120)$\\
   \hline
\end{tabular}
}
\label{tab:node}
\end{table*}

Tables~\ref{tab:time} and \ref{tab:node} show the average computation time (in seconds) and the average number of processed nodes of the algorithms for each class of instances, respectively.
The numbers in parentheses show the number of instances optimally solved within the time limit.
If an algorithm could not solve an instance optimally within the time limit, then we set the computation time to 3600 seconds.
The best computation time among the compared algorithms is highlighted in bold.
The bottom row shows average factors normalized so that the value of BB-ICG+ is set  to one.

We observe that BB-ICG+ attained optimal solutions for all instances within the time limit and ICG, BB-ICG and BB-ICG+ attained best results for 8, 2 and 14 classes, respectively.
We also observed that BB-ICG and BB-ICG+ attained optimal solutions while processing much smaller numbers of nodes than $\textnormal{A}^{\ast}$-MOD and $\textnormal{A}^{\ast}$-DOM.
These results show that ICG attained much better upper bounds than the heuristic functions $h_{mod}$ and $h_{dom}$ for most instances and then the proposed algorithms succeeded in drastically reducing the computation time.
We also observe that ICG attained better results than CG for all classes and BB-ICG+ attained better results than BB-CG for 19 classes.
These results show that the proposed methods improved the performance of the constraint generation algorithm and the branch-and-bound algorithm, respectively.

\section{Conclusion}
In this paper, we present an efficient branch-and-bound algorithm for the non-decreasing submodular function maximization problem based on a BIP formulation with a huge number of constraints.
We propose an improved constraint generation algorithm that starts from a small subset of constraints taken from the constraints and repeats solving a reduced BIP problem and adding a promising set of constraints at each iteration.
We incorporate it into a branch-and-bound algorithm to attain good upper bounds with much less computational effort.
We also introduce a heuristic function and a local search algorithm to attain other good upper and lower bounds quickly.
According to computational results for well-known benchmark instances, our algorithm achieved better performance than the state-of-the-art $\textnormal{A}^{\ast}$~search algorithms and the conventional constraint generation algorithm.

\bibliographystyle{plainnat}
\bibliography{preprint}

\begin{thebibliography}{19}
\providecommand{\natexlab}[1]{#1}
\providecommand{\url}[1]{\texttt{#1}}
\expandafter\ifx\csname urlstyle\endcsname\relax
  \providecommand{\doi}[1]{doi: #1}\else
  \providecommand{\doi}{doi: \begingroup \urlstyle{rm}\Url}\fi

\bibitem[Chen et~al.(2015)Chen, Chen, and Weinberger]{Chen2015}
W.~Chen, Y.~Chen, and K.~Weinberger.
\newblock Filtered search for submodular maximization with controllable
  approximation bounds.
\newblock In \emph{Proceedings of the 18th International Conference on
  Artificial Intelligence and Statistics (AISTATS'15)}, pages 156--164, 2015.

\bibitem[Chvatal(1983)]{Chvatal1983}
V.~Chvatal.
\newblock \emph{Linear Programming}.
\newblock W.~H.~Freeman and Company, New York and Oxford, 1983.

\bibitem[Das and Kempe(2011)]{Das2011}
A.~Das and D.~Kempe.
\newblock Submodular meets spectral: Greedy algorithms for subset selection,
  sparse approximation and dictionary selection.
\newblock In \emph{Proceedings of the 28th International Conference on Machine
  Learning (ICML'11)}, pages 1057--1064, 2011.

\bibitem[Golovin and A.~Krause(2011)]{Golovin2011}
D.~Golovin and A.~A.~Krause.
\newblock Adaptive submodularity: Theory and applications in active learning
  and stochastic optimization.
\newblock \emph{Journal of Artificial Intelligence Research}, 42:\penalty0
  427--486, 2011.

\bibitem[John et~al.(1994)John, Kohavi, and Pfleger]{John1994}
G.~H. John, R.~Kohavi, and K.~Pfleger.
\newblock Irrelevant feature and the subset selection problem.
\newblock In \emph{Proceedings of the 11th International Conference on Machine
  Learning (ICML'94)}, pages 121--129, 1994.

\bibitem[Jovic et~al.(2015)Jovic, Brkic, and Bogunovic]{Jovic2015}
A.~Jovic, K.~Brkic, and N.~Bogunovic.
\newblock A review of feature selection methods with applications.
\newblock In \emph{Proceedings of 38th International Convention on Information
  and Communication Technology, Electronics and Microelectronics (MIPRO'15)},
  pages 1200--1205, 2015.

\bibitem[Kawahara et~al.(2009)Kawahara, Nagano, Tsuda, and
  Bilmes]{Kawahara2009}
Y.~Kawahara, K.~Nagano, K.~Tsuda, and J.~A. Bilmes.
\newblock Submodularity cuts and applications.
\newblock In \emph{Advances in Neural Information Processing Systems 26}, pages
  916--924, 2009.

\bibitem[Kempe et~al.(2003)Kempe, Kleinberg, and Tardos]{Kempe2003}
D.~Kempe, J.~Kleinberg, and E.~Tardos.
\newblock Maximizing the spread of influence through a social network.
\newblock In \emph{Proceedings of the 9th ACM SIGKDD International Conference
  on Knowledge Discovery and Data Mining (KDD'03)}, pages 137--146, 2003.

\bibitem[Kratica et~al.(2001)Kratica, Tosic, Filipovic, Ljubic, and
  Tolla]{Kratica2001}
J.~Kratica, D.~Tosic, V.~Filipovic, I.~Ljubic, and P.~Tolla.
\newblock Solving the simple plant location problem by genetic algorithm.
\newblock \emph{RAIRO-Operations Research}, 35\penalty0 (1):\penalty0 127--142,
  2001.

\bibitem[Krause and Golovin(2014)]{Krause2014}
A.~Krause and D.~Golovin.
\newblock Submodular function maximization.
\newblock In \emph{Tractability: Practical Approaches to Hard Problems}, pages
  71--104. Cambridge University Press, 2014.

\bibitem[Lee et~al.(2009)Lee, Mirrokni, Nagarajan, and Sviridenko]{Lee2009}
J.~Lee, V.~S. Mirrokni, V.~Nagarajan, and M.~Sviridenko.
\newblock Non-monotone submodular maximization under matroid and knapsack
  constraints.
\newblock In \emph{Proceedings of the 41st Annual ACM Symposium on Theory of
  Computing (STOC'09)}, pages 323--332, 2009.

\bibitem[Lovasz(1983)]{Lovasz1983}
L.~Lovasz.
\newblock Submodular functions and convexity.
\newblock In A.~Bachem, M.~Grotschel, and B.~Korte, editors, \emph{Mathematical
  Programming --- The State of the Art}, pages 235--257. Springer, Berlin,
  Heidelberg, 1983.

\bibitem[Minoux(1978)]{Minoux1978}
M.~Minoux.
\newblock Accelerated greedy algorithms for maximizing submodular set
  functions.
\newblock \emph{Lecture Notes in Control and Information Sciences}, 7:\penalty0
  234--243, 1978.

\bibitem[Nemhauser and Wolsey(1981)]{Nemhauser1981}
G.~L. Nemhauser and L.~Wolsey.
\newblock Maximizing submodular set functions: Formulations and analysis of
  algorithms.
\newblock \emph{Studies on Graphs and Discrete Programming}, 11:\penalty0
  279--301, 1981.

\bibitem[Nemhauser et~al.(1978)Nemhauser, Wolsey, and Fisher]{Nemhauser1978}
G.~L. Nemhauser, L.~A. Wolsey, and M.~L. Fisher.
\newblock An analysis of approximations for maximizing submodular set functions
  {I}.
\newblock \emph{Mathematical Programming}, 14\penalty0 (1):\penalty0 265--294,
  1978.

\bibitem[Sakaue and Ishihata(2018)]{Sakaue2018}
S.~Sakaue and M.~Ishihata.
\newblock Accelerated best-first search with upper-bound computation for
  submodular function maximization.
\newblock In \emph{Proceedings of the 32nd AAAI Conference on Artificial
  Intelligence (AAAI'18)}, pages 1413--1421, 2018.

\bibitem[Tuy(1964)]{Tuy1964}
H.~Tuy.
\newblock Concave programming under linear constraints.
\newblock \emph{Soviet Mathematics Doklady}, 5:\penalty0 1437--1440, 1964.

\bibitem[Yu and Liu(2004)]{Yu2004}
L.~Yu and H.~Liu.
\newblock Efficient feature selection via analysis of relevance and redundancy.
\newblock \emph{Journal of machine learning research}, 5:\penalty0 1205--1224,
  2004.

\bibitem[Zeng et~al.(2015)Zeng, Chen, Cao, Qin, Cavazza, and Xiang]{Zeng2015}
Y.~Zeng, X.~Chen, X.~Cao, S.~Qin, M.~Cavazza, and Y.~Xiang.
\newblock Optimal route search with the coverage of users’ preferences.
\newblock In \emph{Proceedings of the Twenty-Fourth International Joint
  Conference on Artificial Intelligence (IJCAI 2015)}, pages 2118--2124, 2015.

\end{thebibliography}
\end{document}